\documentclass[conference]{IEEEtran}
\IEEEoverridecommandlockouts
\usepackage{cite}
\usepackage{amsmath,amssymb,amsfonts}
\usepackage{algorithmic}
\usepackage[dvipdfmx]{graphicx}
\usepackage{textcomp}
\usepackage{xcolor}
\usepackage{bm}
\usepackage{algorithmic}
\usepackage{algorithm}
\usepackage{amsfonts}
\def\BibTeX{{\rm B\kern-.05em{\sc i\kern-.025em b}\kern-.08em
    T\kern-.1667em\lower.7ex\hbox{E}\kern-.125emX}}
\begin{document}

\title{
Security Impact Analysis of Degree of Field Extension in Lattice Attacks on Ring-LWE Problem
}


\author{\IEEEauthorblockN{1\textsuperscript{st} Yuri Lucas Direbieski}
\IEEEauthorblockA{\textit{Grad. Sch. of Sci. and Tech. for Innov.} \\
\textit{Tokushima University}\\
Tokushima, Japan \\
c612235001@tokushima-u.ac.jp}
\and
\IEEEauthorblockN{2\textsuperscript{nd} Hiroki Tanioka}
\IEEEauthorblockA{\textit{Center for Adm. of Info. Tech.} \\
\textit{Tokushima University}\\
Tokushima, Japan \\
https://orcid.org/0000-0003-3404-0855}
\and
\IEEEauthorblockN{3\textsuperscript{rd} Kenji Matsuura\hspace{20pt}}
\IEEEauthorblockA{\textit{Center for Adm. of Info. Tech.} \hspace{20pt}\\
\textit{Tokushima University}\hspace{20pt}\\
Tokushima, Japan \hspace{20pt}\\
ma2@tokushima-u.ac.jp\hspace{20pt}}
\and
\IEEEauthorblockN{\hspace{25pt}4\textsuperscript{th} Hironori Takeuchi}
\IEEEauthorblockA{\hspace{25pt}\textit{Center for Adm. of Info. Tech.} \\
\textit{\hspace{25pt}Tokushima University}\\
\hspace{25pt}Tokushima, Japan \\
\hspace{25pt}takeuchi.hironori@tokushima-u.ac.jp}
\and
\IEEEauthorblockN{\hspace{25pt}5\textsuperscript{th} Masahiko Sano}
\IEEEauthorblockA{\hspace{25pt}\textit{Center for Adm. of Info. Tech.} \\
\hspace{25pt}\textit{Tokushima University}\\
\hspace{25pt}Tokushima, Japan \\
\hspace{25pt}sano@tokushima-u.ac.jp}
\and
\IEEEauthorblockN{\hspace{15pt}6\textsuperscript{th} Tetsushi Ueta}
\IEEEauthorblockA{\hspace{15pt}\textit{Center for Adm. of Info. Tech.} \\
\hspace{15pt}\textit{Tokushima University}\\
\hspace{15pt}Tokushima, Japan \\
\hspace{15pt}https://orcid.org/0000-0001-5810-437X}
}

\maketitle

\begin{abstract}
Modern information communications use cryptography to keep the contents of communications confidential.
RSA (Rivest–Shamir–Adleman) cryptography and elliptic curve cryptography, which are public-key cryptosystems, are widely used cryptographic schemes.
However, it is known that these cryptographic schemes can be deciphered in a very short time by Shor's algorithm when a quantum computer is put into practical use.
Therefore, several methods have been proposed for quantum computer-resistant cryptosystems that cannot be cracked even by a quantum computer.
A simple implementation of LWE-based lattice cryptography based on the LWE (Learning With Errors) problem requires a key length of $O(n^{2})$ to ensure the same level of security as existing public-key cryptography schemes such as RSA and elliptic curve cryptography.
In this paper, we attacked the Ring-LWE (RLWE) scheme, which can be implemented with a short key length, with a modified LLL (Lenstra-Lenstra-Lov\'{a}sz) basis reduction algorithm and investigated the trend in the degree of field extension required to generate a secure and small key.
Results showed that the lattice-based cryptography may be strengthened by employing Cullen or Mersenne prime numbers as the degree of field extension.
\end{abstract}

\begin{IEEEkeywords}
lattice cipher, ring-LWE,  LLL basis reduction
\end{IEEEkeywords}

\section{Introduction}
Modern information communications use cryptography to keep the contents of communications confidential.
The widely used cryptographic schemes include RSA (Rivest–Shamir–Adleman) and elliptic curve cryptography, which are public-key cryptographic schemes.
These cryptographic schemes are based on the fact that current computers cannot solve prime factorization and discrete logarithm problems in a realistic amount of time.
However, these cryptographic schemes are not secure against quantum computer attacks.
It is known that these cryptographic schemes can be cracked in a very short time by Shor's algorithm when quantum computers are put into practical use.
If quantum computers are put into practical use, the security of the public key cryptography currently in use will not be guaranteed.
Therefore, it is desirable to develop and implement post-quantum cryptography that cannot be deciphered even by a quantum computer.
Several methods have been proposed for post-quantum cryptography.
Multivariate polynomial cryptosystems, code-based cryptosystems, lattice cryptosystems, and homomorphic mapping cryptosystems have been widely studied as promising quantum computer cryptosystems.

In this study, we focus on lattice cryptography.
Lattice cryptography is a public-key cryptographic scheme that uses a mathematical problem called the lattice point search problem.
Several implementations have been proposed, most of which are based on the CVP (Closest Vector Problem), SVP (Shortest Vector Problem), and LWE (Learning With Errors).
Our experiments were conducted on lattice ciphers based on LWE in particular, which are described in the following flow.
Sec.~\ref{sec:researches} explains related researches on lattice cryptography based on the LWE and Ring-LWE (RLWE) problem.
Sec.~\ref{sec:implementation} describes the concepts and definitions of lattice basis reduction and the RLWE problem, as well as the attack methods, Kannan's embedding method~\cite{10.5555/2875343.2875346} and the LLL (Lenstra-Lenstra-Lov\'{a}sz) basis reduction algorithm~\cite{1361699995338564224}.
Sec.~\ref{sec:experiment} explains the structure of RLWE problem used in the experiment, and also describe our proposed method and experimental results. 
Finally, section~\ref{sec:conclusion} provides a summary of the study. 

\section{Related Works}\label{sec:researches}
In this section, we first introduce related works on lattice cryptography for the RLWE problem.
Next, a study of an attack on the RLWE problem using Kannan's embedding method and another study of an attack on the RLWE problem using Babai's nearest-neighbor plane algorithm~\cite{10.1007/BFb0023990} are described.

\subsection{Ring-LWE problem}\label{sec:lwe_rlwe}
Lyubashevsky et al. pointed out that the comventional LWE lattice cryptography incur an overhead of $O(n^2)$ in key sharing; therefore, they proposed more efficient LWE-based cryptography, RLWE, by introducing an algebraic structure for the LWE problem~\cite{10.1007/978-3-642-13190-5_1}.
In the RLWE problem under attack,  two RLWE samples are generated with a common secret key, and are used as input for the basis reduction algorithm utilized in the lattice attack. 

\subsection{Safety Analysis of Ring-LWE}\label{sec:safety}
Uesugi et al. experimentally analyze the safety of RLWE by performing a lattice attack using the Progressive-BKZ algorithm on the RLWE problem on a subfield of the cyclotomic field~\cite{isec2021-71}~\cite{isec2022-54}.
%
Here, Kannan's embedding method is used in a lattice attack to extract errors by attributing the CVP to a unique SVP, and the attack is successful if the errors are found correctly.
Their experimental results showed that the squared cyclotomic field is the safest, and that the success rate of lattice attacks decreases and execution time increases when the squared cyclotomic field and basis geometry are different. 

\subsection{Lattice attacks on Ring-LWE field}\label{sec:lattice_attack}
Terada et al.~\cite{8664308} experimentally analyzed the RLWE problem in the decomposition field proposed by Arita et al.~\cite{ARITA20202019CIP0027} and performed a lattice attack on the RLWE problem of some prime cyclotomic fields, comparing execution time, success rate, and root Hermite factor.
Experimental results show that Kannan's embedding method is much faster than Babai's nearest plane algorithm.
In addition, The behavior of the basis reduction algorithm is highly dependent on the structure of the input lattice. 
%
%
As a lattice attack using machine learning, SALSA~\cite{wenger2022salsa} was proposed as a method using transformers to perform modular arithmetic and combine half-trained models with statistical cryptanalysis techniques.
%
Then, it is now clear that SALSA can break LWE problem of medium dimension (up to $n = 128$) with sparse binary secrets. 

\subsection{Summary of related works}
\label{sec:research_summary}
A lattice-based cryptography on the LWE problem was proposed as an implementation of a lattice-based cryptography, however, it is inefficient because it requires the sharing of $O(n^2)$ sized matrix for parameter pre-sharing. 
The more efficient lattice-based cryptography, the RLWE, was proposed by restricting the LWE problem to polynomial rings, where the shared elements are vectors of polynomial coefficients. 
For those RLWE problem-based lattice cryptography security has not been completely proven and there may be instances that provide weak RLWE problem. 
This study focuses on the degree of field extension, which is the dimension of the lattice of the RLWE problem, and conducts a study on the security against lattice attacks using Kannan's embedding method.
%
Generally, a value of a power of two is chosen for the degree of field extension, and a value of about a power of two is considered effective against lattice attacks. However, a related study that conducted safety analysis using the Progressive-BKZ algorithm for the RLWE problem on real and imaginary segments reported that the success rate tends to be low when the dimension is prime number, and the analysis on the number of dimensions is not sufficient.
Therefore, the objective of this study is to examine the effect of the order of the field extension on the lattice attack by constructing an RLWE problem where the order is chosen to the power of two.

\section{Implementation of a safety analysis of the Ring-LWE problem}\label{sec:implementation}

This section explains the lattice basis reduction algorithm and the concept of the RLWE problem, followed by a description of the attack methods, Kannan's embedding method and the LLL basis reduction algorithm.

\subsection{Lattice basis reduction algorithm}\label{sec:lattice}
We explain the CVP as a basic computationally hard problem in lattice. 
CVP is the problem of finding a vector $\bm{x}$ such that, given a lattice $\mathcal{L}$ and target vector $\bm{t}$, $||\bm{t} - \bm{x}|| \leq ||\bm{t} - \bm{y}||$ for all vectors $\bm{y} \in \mathcal{L}$ in CVP$(\mathcal{L}, \bm{t})$. 
Furthermore, there exists an approximate CVP$(\mathcal{L}, \bm{t}, \gamma)$ that introduces a real number $\gamma > 1$. 
Approximate CVP is the problem of finding a vector $\bm{x}$ such that 
$||\bm{t} - \bm{x}|| \leq \gamma||\bm{t} - \bm{y}||$ for all vectors $\bm{y} \in \mathcal{L}$. 
The Babai's nearest plane algorithm~\cite{10.1007/BFb0023990} and Kannan's embedding method~\cite{10.5555/2875343.2875346}, both of which are general-purpose methods for solving LWE problem, are algorithms for solving approximate CVPs\cite{10.1007/978-3-319-71667-1_2}. 
The analysis of lattice cryptography comes down to the problem of computing lattices, including approximate CVPs.
Since many of the lattice point search problems that make up lattice cryptography become more difficult as the rank of the lattice basis increases, the quality of the aforementioned algorithms depends on the rank of the input lattice.
In solving lattice problems such as those described above, a lattice basis reduction algorithm is usually used that makes the lattice basis nearly orthogonal, and in practice, a LLL basis reduction is used as input for Kannan's embedding method.
The root Hermite coefficient~\cite{10.1007/978-3-540-78967-3_3} used to evaluate the quality of the basis reduction algorithm is used as a measure of how well the lattice reduction algorithm shortens the lattice elements.

\subsection{Ring learning with errors problem}\label{sec:errors_problem}
We analyze the security for the RLWE problem on a subfield on a cyclotomic field and a splitting field using the method explained in \cite{isec2021-71}.
Let $K$ the degree of a field extension and $O_K$ be an algebraic number field. Let $\chi _{secret}$ be a secret key probability distribution on $O_K$, and $\chi _{error}$ be an error key probability distribution on $O_K$. Additionally, let $p$ be an integer as polynomial ring, and a field extension $O_K/p O_K$ be denoted as $O_{K, p}$. 
For a probability distribution $\chi$ on a set $X$, write $a \leftarrow \chi$ if $a \in X$ is chosen according to $\chi$, and $U(X)$ is a uniform distribution on $X$.
The RLWE probability distribution such that $O_{K, p}\times O_{K, p}$ is denoted by $R_{K, p, \chi_{secret}, \chi_{error}}$. 
Various parameters $(a, as+e)$ are defined according to these probability distributions, where $a \leftarrow U(O_{K, p}), s \leftarrow \chi_{secret}, e \leftarrow \chi_{error}$ respectively. 
The RLWE problem has two variant. 
One is the problem of distinguishing $R_{K, p, \chi_{secret}, \chi_{error}}$ from $U(O_{K, p} \times O_{K, p})$, which is called the decision RLWE problem. 
The other problem is to find $s \in O_{K, p}$ for any given number of samples $(a_i, a_i s + e_i) \in O_{K, p} \times O_{K, p}$ which chosen from $R_{K, p, \chi_{secret}, \chi_{error}}$. 
This problem is called the search RLWE problem. In public-key cryptography, the problem of finding the secret key corresponds to the search RLWE problem~\cite{10.1007/978-3-642-22792-9_29, 10.1145/2090236.2090262}. 

\subsection{Formalization of Ring-LWE problem}
\label{sec:formalization}
The RLWE problem used in our study is described in more detail below.
The various parameters of the RLWE problem for safety analysis~\cite{10.1007/978-3-642-22792-9_29} are as follows. 
\begin{itemize}
  \item Prime number $q$ $\dots$ a law of each coefficient.
  \item Number of terms in the polynomial $n$ $\dots$ the dimension $K$ of the RLWE problem.
  \item Cyclotomic integers $a$ $\dots$ coefficients chosen from $\chi$.
  \item Secret cyclotomic integer $s$ $\dots$ coefficients take $0$ or $1$.
  \item Error cyclotomic integer $e$ $\dots$ coefficients sufficiently smaller than modulo $q$.
  \item Public key parameter $b$ $\dots$ defined for $a$, $s$ and $e$.
\end{itemize}
Consider an algebraic number field $O_K$ with respect to the degree of the field extension $K$ and define a prime number $q$.
In this case, a prime number $p$ is provided for the polynomial ring $O_{K, q}$.
The parameter $a$ is chosen in the range $(-\frac{q}{2}, \frac{q}{2}]$ 
and according to a probability distribution $\chi$. 
In addition, the parameter $s$ takes the values $0$ or $1$, which is chosen from the probability distribution $\chi _{secret}$. 
For an additional error $e$, define a normal distribution with mean $0$ and variance $\sigma ^2$ for $\sigma > 0$, and sample real numbers from $\chi _{error} = N(0, \sigma ^2)$ to obtain the error distribution.
Moreover, rounded to the nearest integer, each coefficient of the error cyclotomic integer $e$. 
\begin{equation}
  a \leftarrow \chi, \,\,\,\, s \leftarrow \chi _{secret}, \,\,\,\, e \leftarrow \chi _{error}
\end{equation}
From the definition of the RLWE problem, the parameter $b$ is defined for these parameters as
\begin{equation}
  b = a s + e. 
\end{equation}
To implement Kannan's embedding method, a matrix $A$ is constructed using several cyclotomic polynomials $a$.
Where a integral basis of $O_{K, q}$ is $\{x_0, x_1, \dots , x_{n-1}\}$, the parameters $a, s, e$ can be expressed as 
\begin{eqnarray}
  a = \sum_{i=0}^{n-1}a_ix_i , (a_i \in \mathbb{Z}_q), \\
  s = \sum_{i=0}^{n-1}s_ix_i , (s_i \in \mathbb{Z}_q), \\
  e = \sum_{i=0}^{n-1}e_ix_i , (e_i \in \mathbb{Z}_q). 
\end{eqnarray}
However, $\mathbb{Z}_q$ is an Abelian group. 
Additionally, there exists $a_{i, j}$ such that 
\begin{equation}
  ax_j = \sum_{i=0}^{n-1} a_{i, j}x_i , (a_{i, j} \in \mathbb{Z}_q)
\end{equation}
holds for $0 \leq j \leq n-1$. 
Using $ax_j$, it can be written as 
\begin{eqnarray}
  as &= &\sum_{j=0}^{n-1} s_j(ax_j),  \nonumber \\
  &= &\sum_{j=0}^{n-1} s_j(\sum_{i=0}^{n-1} a_{i, j}x_i),  \nonumber \\
  &= &\sum_{i=0}^{n-1} (\sum_{j=0}^{n-1} a_{i, j}s_j)x_i. 
\end{eqnarray}
Here, by setting 
\begin{eqnarray}
  \bm{b} = (b_0, \dots , b_{n-1})^T, \\
  \bm{s} = (s_0, \dots , s_{n-1})^T, \\
  \bm{e} = (e_0, \dots , e_{n-1})^T, 
\end{eqnarray}
and then we obtain the following equation, 
\begin{equation}
  \label{eq:RLWE}
  \bm{b} = A\bm{s} + \bm{e}  \, (\mathrm{mod} \, q), 
\end{equation}
where $A$ is
\begin{equation}
  A = 
  \begin{pmatrix}
    a_{0, 0} & \dots & a_{0, n-1} \\
    \vdots & \ddots & \vdots \\
    a_{n-1, 0} & \dots & a_{n-1, n-1}
  \end{pmatrix}. 
\end{equation}
Similarly, given that another RLWE sample $(A', \bm{b}')$ is generated using the common $\bm{s}$, there is $A' \in (\mathbb{Z}/q\mathbb{Z})^{n \times n}, \bm{b}' (\in \mathbb{Z}^n), \bm{e}' (\in \mathbb{Z})^n$ satisfying 
\begin{equation}
  \bm{b}' = A'\bm{s} + \bm{e}'  \, (\mathrm{mod} \, q). 
\end{equation}
Therefore, it can be expressed as 
\begin{equation}
  \label{eq:2RLWE}
  \begin{pmatrix}
    \bm{b} \\
    \bm{b}'
  \end{pmatrix}
  = 
  \begin{pmatrix}
    A \\
    A'
  \end{pmatrix}
  \cdot s
  +
  \begin{pmatrix}
    \bm{e} \\
    \bm{e}'
  \end{pmatrix}
   \, (\mathrm{mod} \, q). 
\end{equation}
This corresponds to the RLWE problem with dimension $n$ and number of samples $2n$. 
Consider the case where an attack is made on the RLWE problem such that the equation is satisfied.

%
%

\subsection{Kannan's embedding}\label{sec:kannan}
Apply Kannan's embedding method to the RLWE sample matrix $A$ obtained for the RLWE problem described in Sec.~\ref{sec:errors_problem}.
Kannan's embedding method~\cite{10.5555/2875343.2875346} that is a generic solution for CVP, extracts errors as a difference vector by attributing the CVP to a unique SVP~\cite{10.1007/978-3-319-71667-1_2}. 
Consider the basis of the $n$-dimensional lattice $\mathcal{L} \subseteq \mathbb{Z}^n$ and the solution vector $\bm{v} = \sum_{i=1}^{n}v_i\bm{b}_i \in \mathcal{L}, (\exists v_i \in \mathbb{Z})$ of the CVP for the target vector $w$. 
In this case, assuming that the norm $||\bm{e}||$ of the difference vector $\bm{e} = \bm{w} - \bm{v}$ between the target vector and  of the solution vector are sufficiently small.
Kannan's embedding method constructs a new lattice $\bar{\mathcal{L}}$ containing this difference vector as the shortest vector. 
The lattice $\bar{\mathcal{L}} \in \mathbb{Z}^{n+1}$ consists of a $(n+1)$-dimensional lattice generated by a linearly independent vector $(n+1)$ of vectors $(\bm{b}_1, 0), \dots , (\bm{b}_n, 0), (\bm{w}_1, M) \in \mathbb{Z}^{n+1}$ for a fixed positive constant $M \in \mathbb{Z}$. 
Furthermore, 
\begin{eqnarray}
(e, M) & = & \left(\bm{w} - \sum_{i=1}^{n} v_i\bm{b}_i, M \right) \\
       & = & -v_1(\bm{b}_1, 0) \dots - v_n(\bm{b}_n, 0) + (\bm{w}, M) \nonumber
\end{eqnarray}
where the vector $(\bm{e}, M)$ is contained in the lattice $\bar{\mathcal{L}}$. 

In particular, when vector $(\bm{e}, M)$ is the shortest vector on lattice $\bar{\mathcal{L}}$, by solving for the SVP on the lattice $\bar{\mathcal{L}}$, the difference vector $||\bm{e}||$ can be obtained.

As a result, the solution $\bm{v} = \bm{w} - \bm{e}$ of CVP can be obtained.
Consider applying Kannan's embedding method to the sample RLWE problem described in Sec.~\ref{sec:formalization}.
Using the elements of \eqref{eq:RLWE}, construct the matrix $A_c, \bm{b}_c$ as 
\begin{equation}
  A_c = \begin{pmatrix}
    A & A' \\
    qI_n & O \\
    O & qI_n
  \end{pmatrix}
  ,  \, 
  \bm{b}_c = \begin{pmatrix}
    \bm{b} & \bm{b}'
  \end{pmatrix}. 
\end{equation}
Finding the errors $\bm{e}, \bm{e}'$ is equivalent to solving for the CVP on the lattice, which is generated by the row vectors of $A_c$ with $\bm{b}_c$ as the target vector. 
By solving CVP to find the vector which closest to the target vector, the error can be extracted by difference. 
Argorithm~\ref{arg:Kannan} shows the algorithm of Kannan's embedding technique.
%
  \begin{algorithm}[!t]
\small
      \caption{Kannan's embedding technique}
      \label{arg:Kannan}
      \begin{algorithmic}[1]    
      \REQUIRE Basis $\{\bm{b}_1, \dots , \bm{b}_n\}$ of $n$-dimensional lattice $\mathcal{L} \subseteq \mathbb{Z}^n$, target vector $\bm{w}$
      \ENSURE The lattice vector $\bm{v} \in \mathcal{L}$, or $false$ closest to $\bm{w}$
      \STATE Lattice $\bar{\mathcal{L}} \leftarrow \begin{pmatrix}
        B & \bm{w} \\
        0 & 1
      \end{pmatrix}$
      generated from $B and \bm{w}$
      \STATE Simplify $\bar{\mathcal{L}}$.
      \IF {$\bm{v} \leftarrow \pm (\bm{v}_1, \dots , \bm{v}_n, 1)$}
      \STATE return $\bm{v}$
      \ELSE
      \STATE return $false$
      \ENDIF
      \end{algorithmic}
  \end{algorithm}
%
The operation of converting the basis to a basis of the same lattice $\mathcal{L}$, where each basis vectors are short and nearly orthogonal to each other, when given a basis $\{\bm{b}_1, \dots , \bm{b}_n\}$ of the lattice $\mathcal{L}$. 
It is called lattice basis reduction algorithm. 

\subsection{LLL lattice basis reduction algorithm}\label{sec:LLL}
In the experiment, the LLL lattice basis reduction algorithm~\cite{1361699995338564224} is used as the algorithm to generate the reduced basis with Kannan's embedding method. 
LLL lattice basis reduction algorithm is a well-known and representative algorithm, which is an efficient algorithm to approximate SVP on an $n$-dimensional lattice. 
For basis reduction on $2$-dimensional lattices, Lagrange's basis reduction algorithm can be used to reduced a $2$-dimensional lattice in polynomial time and obtain an exact solution. 
However, in the general dimension case, the LLL basis reduction algorithm is used to approximate the reduction basis. 

In particular, it is called the LLL basis reduction when the following two conditions are satistified.
\begin{enumerate}
  \item $|\mu_{i, j}| \leq \frac{1}{2}$ for natural numbers $1 \leq j < i \leq n$ (Size reduction)
  \item $(\delta - \mu_{i+1, i}^2)|b_i ^*|^2 \leq |b_{i+1}^*|^2$ for natural numbers $1 \leq j < n$ (Lov\'{a}sz's condition)
\end{enumerate}
The LLL basis reduction algorithm reduces the lattice by repeating two operations: first, the basis is reduced to size, and then the Lov\'{a}sz condition is checked if it is satisfied; if not, the process starts over.  
Argorithm~\ref{arg::LLL} shows the LLL basis simplification algorithm.
%
  \begin{algorithm}[!t]
\small
      \caption{LLL basis reduction algorithm}
      \label{arg::LLL}
      \begin{algorithmic}[1]    
      \REQUIRE Basis $\{\bm{b}_1, \dots , \bm{b}_n\}$ of $n$-dimensional lattice $\mathcal{L} \subseteq \mathbb{Z}^n$, parameter $0 \leq \sigma < 1$
      \ENSURE Simplified basis $\{\bm{b}_1, \dots , \bm{b}_n\}$ for $n$-dimensional lattice $\mathcal{L} \subseteq \mathbb{Z}^n$ with respect to $\sigma$
      \STATE Basis $\{\bm{b}_1, \dots , \bm{b}_n\}$ GSO coefficient $(\mu_{i, j})_{1 \leq j < i \leq n} \leftarrow GSO(\bm{b}_1, \dots , \bm{b}_n)$
      \STATE Compute $||b_i ^*||^2$, $(1 \leq i \leq n)$
      \STATE $k = 2$
      \WHILE {$k \leq n$}
      \FOR {$j = k - 1$ downto $1$}
      \IF {$|\mu_{k, j}| > \frac{1}{2}$}
      \STATE $b_k = b_k - \lfloor \mu_{k, j}\rceil b_j$
      \STATE $b_1 ^*, \dots , b_n ^* , (\mu_{i, j})_{1 \leq j < i \leq n} = GSO(b_1, \dots , b_n)$
      \ENDIF
      \ENDFOR
      \IF {$(\delta - \mu_{i+1, i}^2)|b_i ^*|^2 \leq |b_{i+1}^*|^2$}
      \STATE $k = k + 1$
      \ELSE
      \STATE $swap(b_k, b_{k-1})$
      \STATE $b_1 ^*, \dots , b_n ^*, (\mu_{i, j})_{1 \leq j < i \leq n} = GSO(b_1, \dots , b_n)$
      \STATE Compute $||b_i ^*||^2$, $(1 \leq i \leq n)$
      \STATE $k = max\{2, k-1\}$
      \ENDIF
      \ENDWHILE
      \STATE return $\{\bm{b}_1, \dots , \bm{b}_n\}$
      \end{algorithmic}
  \end{algorithm}

To apply Kannan's embedding method, first find the reduced lattice basis of the input.
Therefore, the LLL basis reduction algorithm is performed on matrix $A_c$ to obtain the reduced matrix $A_{LLL}$.
Applying Kannan's embedding method to this matrix $A_{LLL}$ yields a matrix $W \in (O_{K, q})^{(2n+1)\times (2n+1)}$ such that
\begin{equation}
  W = \begin{pmatrix}
    A_{LLL} & 0^T \\
    \bm{b}_c & 1
  \end{pmatrix}. 
\end{equation}
Simplifying the matrix $W$ again using the LLL basis reduction algorithm yields the error vector $(\bm{e}, \bm{e}', 1)$ from the reduced matrix. 
In our experiments, the LLL lattice basis reduction algorithm is used as the algorithm used to generate the reduced basis by Kannan's embedding method.

\section{Proposed method and experiment}\label{sec:experiment}
To analyze the impact of the degree of field extension which is the dimension of the lattice of the RLWE problem on lattice cryptography, this section describes the RLWE problem used in the lattice attack experiments, followed by a description of the experiments and the environment.
For various parameters of the RLWE problem, let the prime number $q = 1997$ and the Gaussian distribution parameter $\sigma = \frac{4}{\sqrt{2\pi}}$\cite{10.1007/978-3-319-03515-4_1}.
The values of the cyclotomic integers $a, s, e$ follow the definition of Sect.~\ref{sec:formalization}.
In this experiment, three integers of powers of 2, $n = \{32, 64, 128\}$, are chosen as reference integers, two integers which are $\pm 1$ and the two larger and smaller primes closest to each reference integer are chosen, and $100$ times decoding experiments are performed for $21$ integers as degree of field extension $d$ to investigate its effect on the lattice attack.
The integers representing the degree of field extension to be handled are $d = \{23, 29, 31, 32, 33, 37, 41\}$ for $n = 32$, $d = \{59, 61, 63, 64, 65, 67, 71\}$ for $n = 64$, and $d = \{109, 113, 127, 128, 129, 131, 137\}$ for $n = 128$.
For the RLWE problem described above, the equations mentioned in Sec.~\ref{sec:formalization} are constructed, Kannan's embedding method mentioned in Sec.~\ref{sec:kannan} is performed, and the error $e_{sought}$ is extracted. 
Equation~\eqref{eq:RLWE} is deformed to obtain the secret-key $s_{sought}$ using \eqref{eq:deform_s} after extracting the error $e_{sought}$. 
\begin{align}
  \bm{b} = &A\bm{s} + \bm{e} \, (\mathrm{mod} \, q) \\
  \label{eq:deform_s}
  \bm{s} = &(\bm{b} - \bm{e}) \cdot A^{-1} \, (\mathrm{mod} \, q)
\end{align}
Finally, the plaintext is decoded using the secret-key $s_{sought}$. 
If the plaintext is equal to the original plaintext, the attack is successful. 

\subsection{Experimental procedure}\label{sec:experiment_procedure}
The following flow of experiments will be conducted.
\begin{enumerate}
\item Generating two RLWE samples including $(A, \bm{b})$, specifically $\bm{b}$ is conducted by $\bm{e}$ which is based on a secret key $\bm{s}$
\item Encoding a plaintext $\bm{m}$ with $(A, \bm{b})$ to a ciphertext $C$
\item Obtaining a error $e_{sought}$ from $(A, \bm{b})$ using Kannan's embedding
\item Finding a secret key $s_{sought}$ with $e_{sought}$
\item Decoding $C$ to a plaintext $\bm{m}_{sought}$ with $s_{sought}$
\item Comparing $\bm{m}_{sought}$ with $\bm{m}$
\end{enumerate}
First, convert the plaintext $m$ from a sequence of n random numbers consisting of $0$ or $1$ values to a cyclotomic integer $\bm{m} = \sum_{i=0}^{n-1} m_i x_i (m_i \in \mathbb{Z}_q)$ and prepare $100$ plaintext sentences $m$.

Second, consider the RLWE problem satisfying \eqref{eq:2RLWE} and the generated plaintext is encrypted using the public-key $(A, \bm{b})$.
Specifically, a small random number $\bm{v} = \sum_{i=0}^{n-1} v_i x_i (v_i \in \mathbb{Z}_q)$, noise $e_0, e_1 \leftarrow (0, \sigma)$ is generated and $\bm{c}_0, \bm{c}_1$ defined by
\begin{align}
  \bm{c}_0 = &\bm{b}v + 2e_0 + \bm{m} \\
  \bm{c}_1 = &A\bm{v} + 2e_1 . 
\end{align}
Thus, ciphertext $C = (\bm{c}_0, \bm{c}_1)$ is obtained.
To decrypt ciphertext $C$, the plaintext $\bm{m}$ is decrypted according to \eqref{eq:decode} using secret key $\bm{s}$~\cite{arita}.
\begin{align}
  \label{eq:decode} 
  \bm{c}_0 - \bm{s} \bm{c}_1 & \equiv \bm{b}\bm{v} + 2e_0 + \bm{m} - \bm{s}(A\bm{v} + 2e_1) \\ \nonumber
  & \equiv \bm{m} + 2(\bm{e} \bm{v} + e_0 - \bm{s}e_1) \, (\mathrm{mod} \, q) 
\end{align}
Hence, Kannan's embedding method is applied to the matrix $A, AA'$ and target vectors $\bm{b}, \bm{b}'$, and the error $e_{sought}$ is extracted according to Sec.~\ref{sec:kannan}. 

The secret key $s_{sought}$ is obtained according to \eqref{eq:deform_s}, and the plaintext $\bm{m}_{sought}$ can be decrypted according to \eqref{eq:decode} with error $\bm{e}_{sought}$ and the attack is successful if the decrypted plaintext $\bm{m}_{sought}$ is equal to the original plaintext $\bm{m}$. 
The decryption time is defined as the time from the start of the execution of Kannan's embedding method to the time when the secret key $\bm{s}_{sought}$ is obtained.
The above experiment is performed $100$ times for each degree of a field extension $n$, and the number of successful attacks and the average execution time are measured. 

\subsection{Experimental conditions}\label{sec:condition}
The experiment was implemented with 
CPU: Intel\textregistered \, Core\texttrademark \, i5-3230M, 2.60GHz $\times$ 4, Memory: 8GB, DDR3, 1600MHz, OS: Ubuntu 20.04.5 LTS, and Libraries: Python 3.8.10, SageMath version 9.0.
%
%

\subsection{Results and discussion}\label{sec:results}
\begin{figure}[!t]
\centerline{\includegraphics[width=0.34\textwidth]{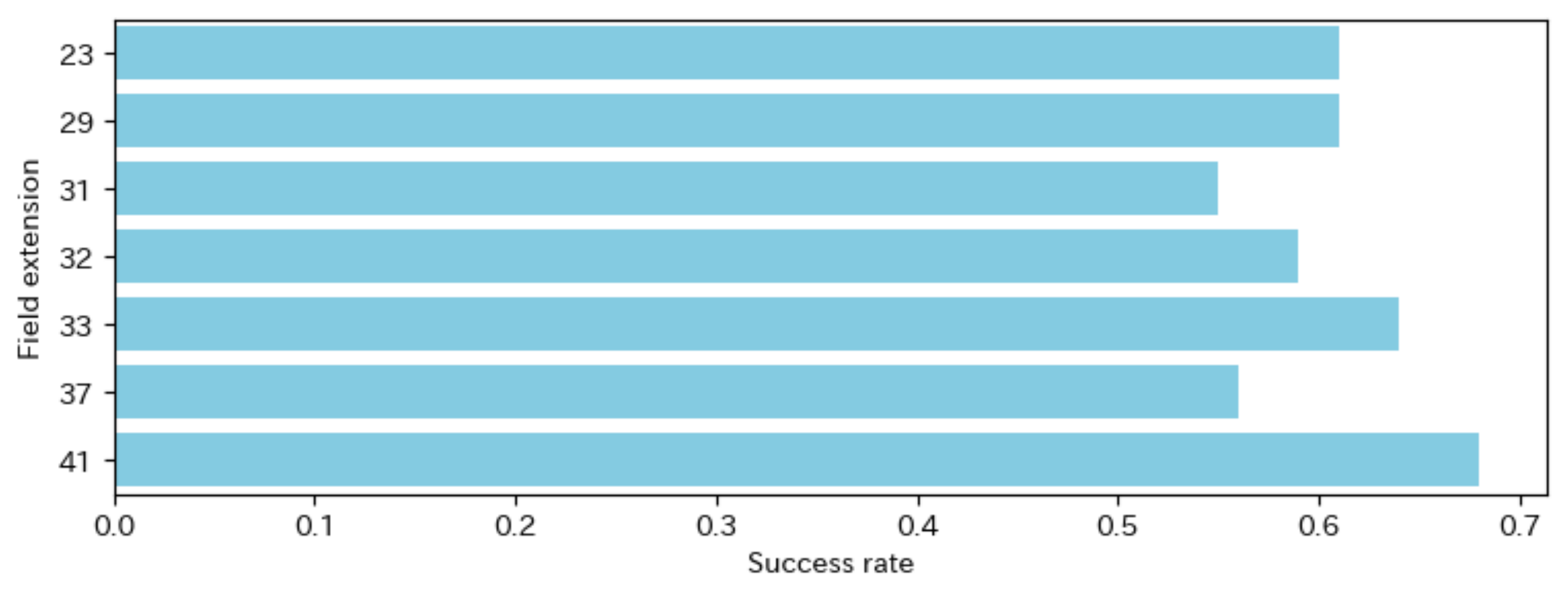}}
%
\centerline{\includegraphics[width=0.34\textwidth]{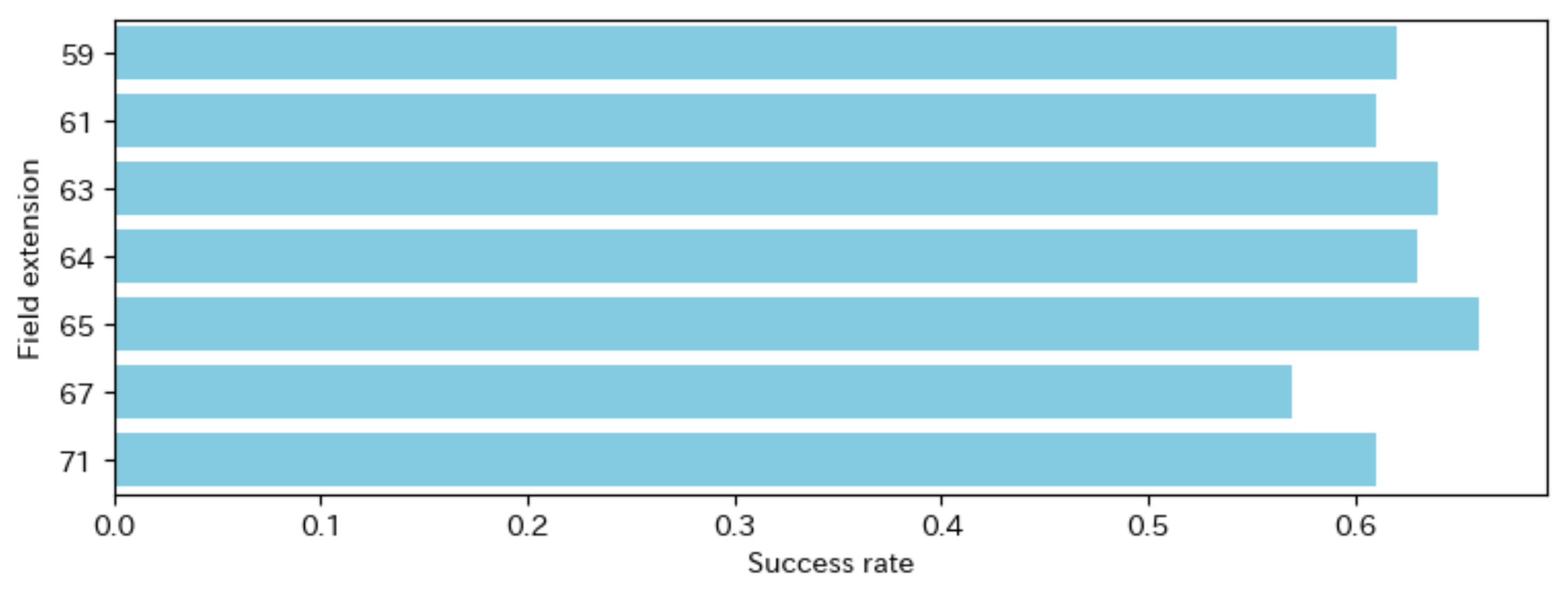}}
%
\centerline{\includegraphics[width=0.34\textwidth]{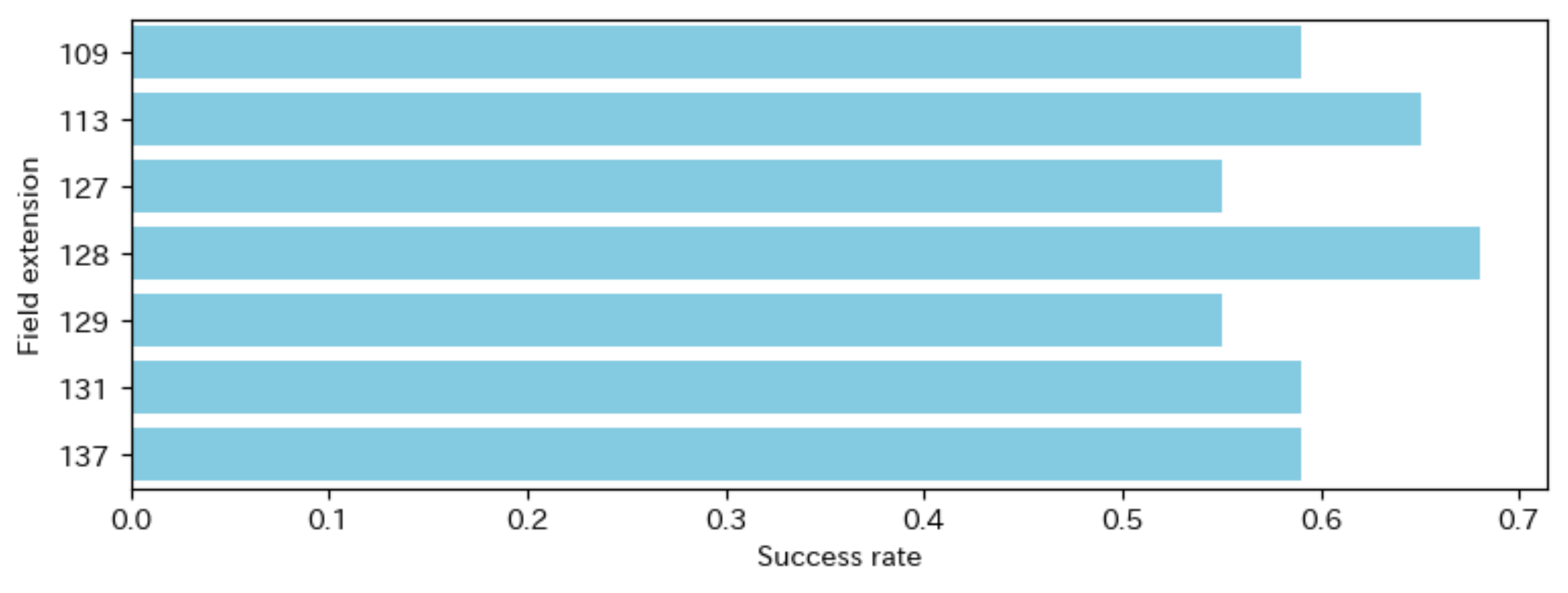}}
\caption{Success rate to field extension around $M_{32}$, $M_{64}$ and $M_{128}$.}
\label{fig:M_32_64_128_rate}
\end{figure}

\begin{figure}[!t]
\centerline{\includegraphics[width=0.34\textwidth]{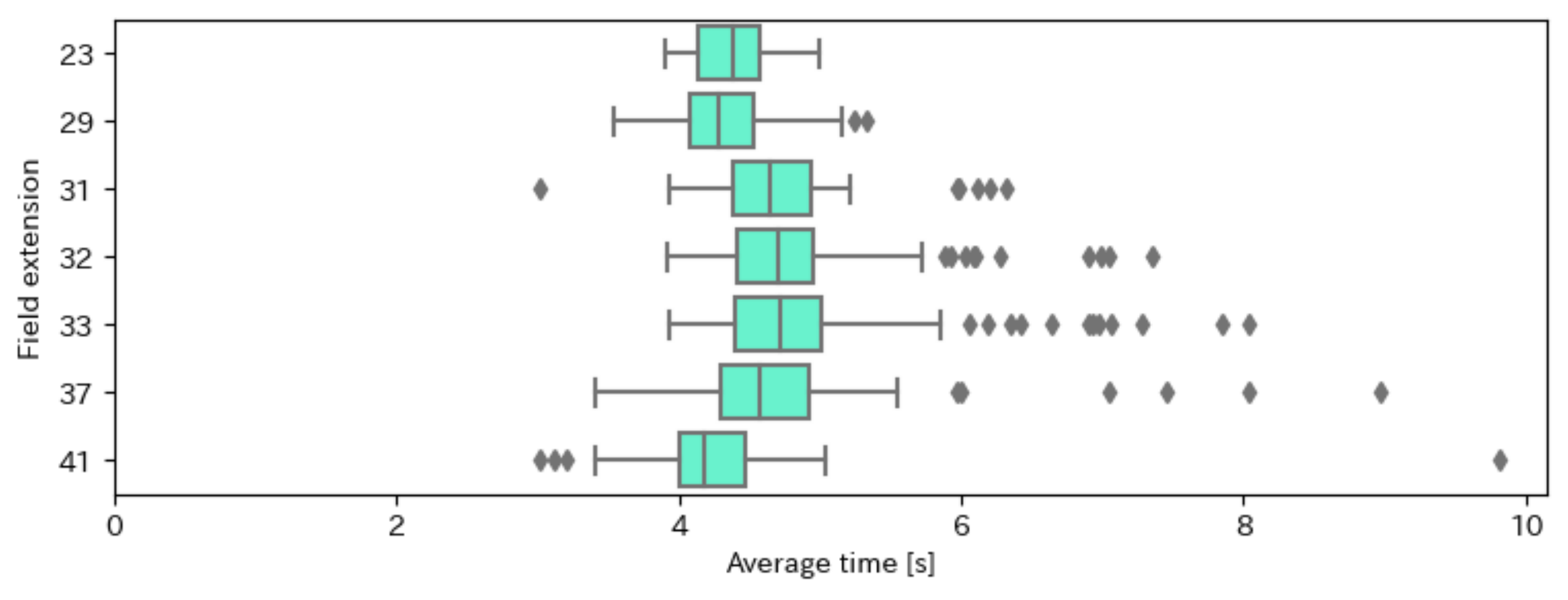}}
%
\centerline{\includegraphics[width=0.34\textwidth]{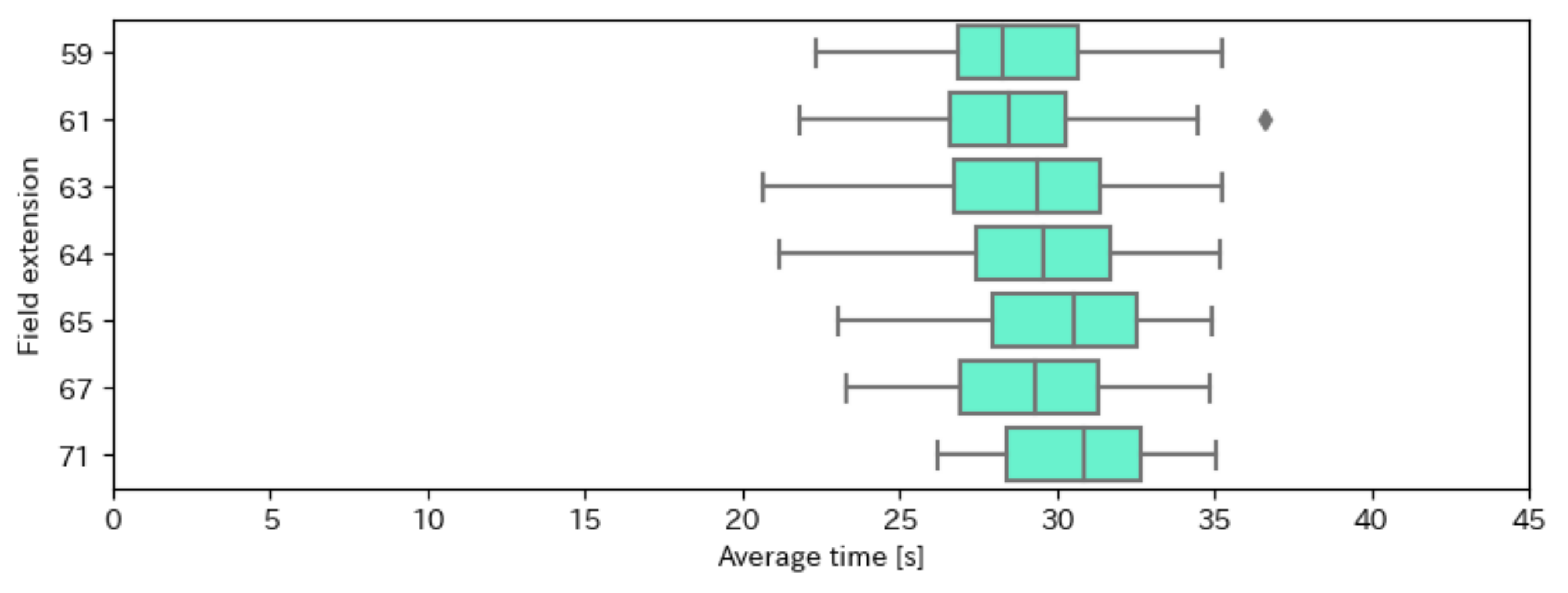}}
%
\centerline{\includegraphics[width=0.34\textwidth]{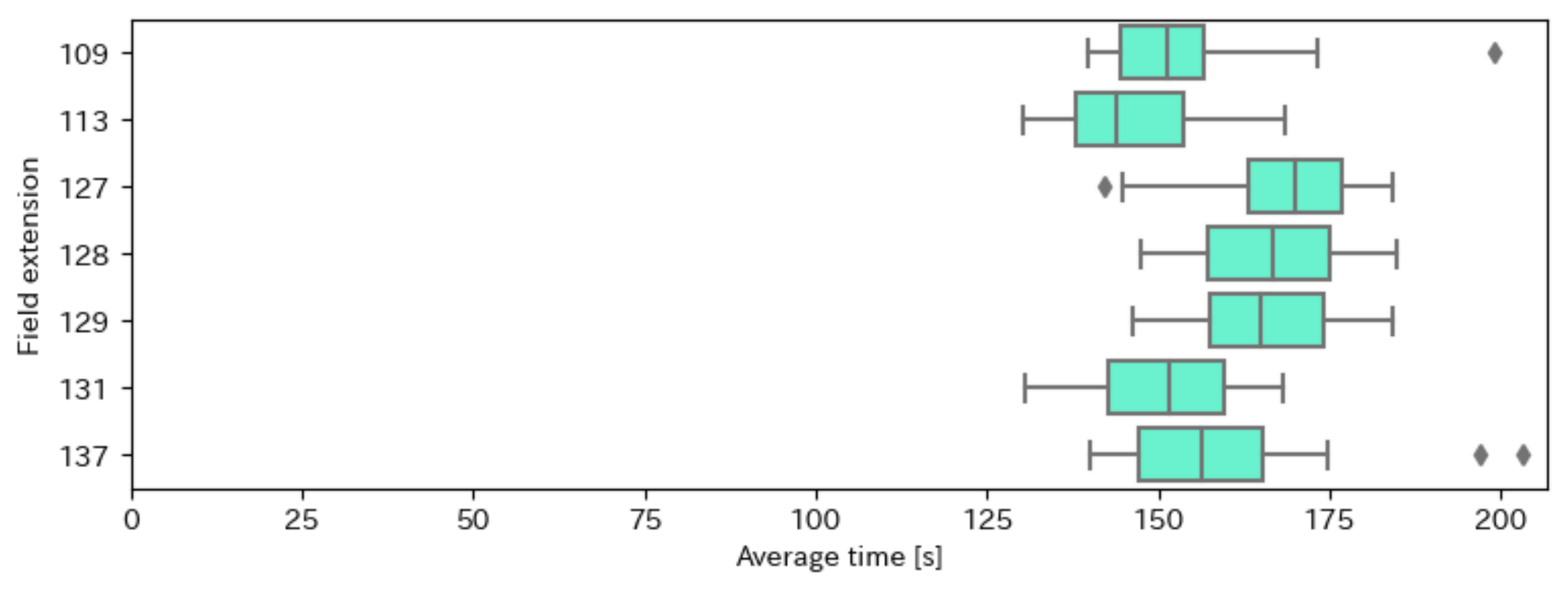}}
\caption{Average time to field extension around $M_{32}$, $M_{64}$ and $M_{128}$.}
\label{fig:M_32_64_128_time}
\end{figure}

The results of the experiment are presented and a discussion of the experimental results is provided.
The results are shown in Table~\ref{tab:m32}, \ref{tab:m64}, and \ref{tab:m128} for $M_{32}, M_{64}, $ and $M_{128}$, respectively. 
\begin{table}[!t]
\caption{Results of attack to field extension around $M_{32}$}
\begin{center}
\scriptsize
\begin{tabular}{cccc}
\hline
\textbf{\textit{Field extension}} & \textbf{\textit{Success rate}} & \textbf{\textit{Average time} [s]} \\ 
\hline
$23$ & $0.61$ & $4.38$ \\
$29$ & $0.61$ & $4.32$ \\
$31$ & $0.55$ & $4.65$ \\
$\mathbf{32}$ & $0.59$ & $4.81$ \\
$33$ & $0.64$ & $4.93$ \\
$37$ & $0.56$ & $4.73$ \\
$41$ & $0.68$ & $4.26$ \\
\hline
\end{tabular}
\label{tab:m32}
\end{center}

\vspace{2mm}
\caption{Results of attack to field extension around $M_{64}$}
\begin{center}
\scriptsize
\begin{tabular}{cccc}
\hline
\textbf{\textit{Field extension}} & \textbf{\textit{Success rate}} & \textbf{\textit{Average time} [s]} \\ 
\hline
$59$ & $0.62$ & $28.83$ \\ 
$61$ & $0.61$ & $28.41$ \\ 
$63$ & $0.64$ & $29.17$ \\ 
$\mathbf{64}$ & $0.63$ & $29.54$ \\ 
$65$ & $0.66$ & $30.20$ \\ 
$67$ & $0.57$ & $29.25$ \\ 
$71$ & $0.61$ & $30.61$ \\ 
\hline
\end{tabular}
\label{tab:m64}
\end{center}
%
\vspace{2mm}
\caption{Results of attack to field extension around $M_{128}$}
\begin{center}
\scriptsize
\begin{tabular}{cccc}
\hline
\textbf{\textit{Field extension}} & \textbf{\textit{Success rate}} & \textbf{\textit{Average time} [s]} \\ 
\hline
$109$ & $0.59$ & $152.04$ \\ 
$113$ & $0.65$ & $145.59$ \\ 
$127$ & $0.55$ & $169.13$ \\ 
$\textbf{128}$ & $0.68$ & $166.51$ \\ 
$129$ & $0.55$ & $165.55$ \\ 
$131$ & $0.59$ & $150.45$ \\ 
$137$ & $0.59$ & $157.05$ \\ 
\hline
\end{tabular}
\label{tab:m128}
\end{center}
\vspace{-2mm}
\end{table}
Fig.~\ref{fig:M_32_64_128_rate} compares the success rates. Fig.~\ref{fig:M_32_64_128_time} compares processing times.
The experimental results showed that the success rate of prime numbers for field extension tended to be lower than that of integers to the power of two for field extension, and that their strength as a cipher tended to be higher. 
The success rates for the prime numbers before and after the reference integers ($32$, $64$ and $128$) and the other integers are compared by Mann-Whitney's U-test ($p > 0.05$), resulting in $p = 9.06\times10^{-3}, r = 0.57$ to be significant differences.
%
On the other hand, the analysis time tended to be longer for integers around the power of two, including integers to the power of two for field extension.
For $M_{32}$ $M_{64}$, and $M_{128}$, respectively, $p = 2.21\times10^{-29}$, $4.08\times10^{-5}$, and $1.60\times10^{-5}$ with the Shapiro-Wilk test ($p > 0.05$), 
and the Kruskal-Wallis test ($p > 0.05$, $n = 6$) with $5.32\times10^{-25}$, $2.55\times10^{-7}$, and $1.29\times10^{-61}$.
Hence, between reference integers ($32$, $64$ and $128$) and the prime numbers before and after the reference integers are confirmed by the Steel-Dwass test ($p > 0.05$): $p_{31} = 0.90$, $p_{37} = 0.88$, $p_{61} = 0.51$, $p_{67} = 0.90$, $p_{127} = 0.49$, $p_{129} = 0.90$ to be no difference.
%
%

\section{Conclusion}\label{sec:conclusion}
In this paper, we performed the lattice reduction algorithm as the RLWE problem with various integers to a field extension and performed a lattice attack by solving a unique SVP using Kannan's embedding method.
Results showed that the attack success rate was lower when prime numbers were used as the degree of field extension, and that the processing time tended to be longer for integers close to powers of two as the degree of field extension, suggesting that the lattice-based cryptography may be strengthened by employing Cullen or Mersenne prime numbers as the degree of field extension.
%
Future work should include comparisons with attacks using other algorithms, such as the BKZ lattice reduction algorithm and machine learning-based algorithms.

\section*{Acknowledgment}
I would like to thank my colleagues for their help in creating and discussing the experimental environment. 
This work was supported by JSPS KAKENHI Grant Number JP22K12293 and JP18K11572. 


\bibliography{lucas} 
\bibliographystyle{IEEEtran} 

\newpage

\copyright 2023 IEEE. Personal use of this material is permitted. Permission from IEEE must be obtained for all other uses, in any current or future media, including reprinting/republishing this material for advertising or promotional purposes, creating new collective works, for resale or redistribution to servers or lists, or reuse of any copyrighted component of this work in other works.

\end{document}